\documentclass[a4paper,10pt]{article}

\usepackage[sans]{dsfont}
\usepackage[mathscr]{euscript}
\usepackage[centertags]{amsmath}
\usepackage{cite}
\usepackage{amsfonts}
\usepackage{amsthm,enumerate,amssymb}
\usepackage{graphicx}
\usepackage{bm,bbm}
\newcommand{\id}{{\mathbbm{1}}}

\usepackage{hyperref}
\usepackage{sidecap}

\newcommand{\rmd}{\mathrm{d}}

\newcommand{\rme}{\mathrm{e}}

\newcommand{\abs}[1]{\left|#1\right|}

\newcommand{\openone}{\id}

\newcommand{\Ao}{\mathsf{A}}
\newcommand{\Bo}{\mathsf{B}}
\newcommand{\Co}{\mathsf{C}}
\newcommand{\Fo}{\mathsf{F}}
\newcommand{\Mo}{\mathsf{M}}
\newcommand{\Xo}{\mathsf{X}}
\newcommand{\Yo}{\mathsf{Y}}
\newcommand{\Zo}{\mathsf{Z}}

\newcommand{\Srel}[2]{S\big(#1\|#2\big)}
\newcommand{\ii}{c}
\newcommand{\Icomp}{\ii_{\rm inc}}

\newcommand{\Div}[2]{D\big(#1\|#2\big)}
\newcommand{\Divm}[2]{\overline{D}\big(#1\|#2\big)}

\newcommand{\sddiv}[2]{S[#1 \| #2]}

\newcommand{\Acal}{\mathcal{A}}

\newcommand{\Tr}{\operatorname{Tr}}

\begin{document}

\title{Uncertainty relations and information loss \\ for spin-1/2 measurements}

\author{A. Barchielli$^{1,2,3}$, M. Gregoratti$^{1,2}$
\\
\\
$^1$ \textsl{Politecnico di Milano, Dipartimento di Matematica}, \\
\textsl{Piazza Leonardo da Vinci 32, I-20133 Milano, Italy}.\\
$^2$ Istituto Nazionale di Alta Matematica (INDAM-GNAMPA).\\
$^3$ Istituto Nazionale di Fisica Nucleare (INFN), Sezione di Milano.}


\maketitle

\begin{abstract}
We formulate entropic measurements uncertainty relations (MURs) for a spin-1/2 system. When incompatible observables are approximatively jointly measured, we use relative entropy to quantify the information lost in approximation and we prove positive lower bounds for such a loss: there is an unavoidable information loss. Firstly we allow only for \emph{covariant} approximate joint measurements and we find state-dependent MURs for two or three orthogonal spin-1/2 components. Secondly we consider \emph{any} possible approximate joint measurement and we find state-independent MURs for two or three spin-1/2 components. In particular we study how MURs depend on the angle between two spin directions. Finally, we extend our approach to infinitely many incompatible observables, namely to the spin components in all possible directions. In every scenario, we always consider also the characterization of the optimal approximate joint measurements.

\bigskip

\end{abstract}

\section{Introduction}\label{intro}
Measurement uncertainty relations (MURs) quantify to which extent one can approximate a set of measurements of incompatible observables by a single joint measurement; for a recent presentation of the modern theory of quantum measurements see, e.g., \cite{BLPY}. Various approaches have been proposed to quantify this approximation, such as distances for probability measures \cite{Wer04,BLW14a,BLW14b,BusLW14,DamSW15,Wer16,BullB18} or conditional entropies \cite{BusHOW,CF15,AbbB16}. Our approach is to see the joint measurement approximation of incompatible observables as a loss of information and to quantify it by the use of the relative entropy \cite{BGT17,BGT18}.

Here we describe the spin-1/2 variant of Ref.\ \cite{BGT18} with some new results (proved in Section \ref{sec6}). We formulate both state-dependent and state-independent entropic MURs for spin components, which are always incompatible when different directions are measured. The state-dependent MURs are new and they arise, as done in \cite{BGT17} for position and momentum, from a selection of the admissible approximate joint measurements based on the covariance properties of the target observables.

We start with two or three incompatible spin components, and then we extend our approach to an approximate joint measurement of infinite observables, the components of the spin along all the possible directions.

Let us remark that the entropic approach naturally selects covariant approximate joint measurments and, in several remarkable cases, gives uniqueness of the optimal one and completely characterises it.

\section{Information lost by approximate measurements}\label{sec2}
Given a spin-1/2 quantum system, we want to quantify the information lost when two \emph{incompatible} sharp observables $\Ao$ and $\Bo$, the spin components along independent unit vectors $\vec a$ and $\vec b$, are approximated by two \emph{compatible} POVM observables $\Mo_{[1]}$ and $\Mo_{[2]}$. For general $\vec a$ and $\vec b$, the corresponding sharp observables are identified by the pvm's
\begin{equation}\label{ABspin}
\Ao(x) = \frac{1}{2}
\left(\id+x\,\vec {a} \cdot \vec {\sigma}\right),  \quad x=\pm1, \qquad
\Bo(y)= \frac{1}{2} \left(\id+ y\,\vec {b} \cdot \vec {\sigma}\right),\quad y=\pm1,
\end{equation}
while $\Mo_{[1]}$ and $\Mo_{[2]}$ are the marginals of some bi-observable $\Mo$
$$
\Mo(x,y)\geq0, \quad x,y=\pm1,\qquad \sum_{x,y}\Mo(x,y)=\id.
$$

Given the system state
\begin{equation}\label{rhor}
\rho=\frac{1}{2}\left(\id+\vec {r} \cdot \vec {\sigma}\right),\qquad r\equiv |\vec r|\leq1,
\end{equation}
if the target observable $\Ao$ is approximated by $\Mo_{[1]}$, then the target probability distribution $\Ao^\rho(x)=\Tr\{\Ao(x)\rho\}$ is approximated by the probability distribution $\Mo_{[1]}^\rho(x)=\Tr\{\Mo_{[1]}(x)\rho\}$. This causes a loss of information that is quantified by the relative entropy $\Srel{\Ao^\rho}{\Mo_{[1]}^\rho}$ \cite[p.\ 51]{BA02}, \cite[Sect.\ 2.3]{CovT06}.

Given two probabilities $p$ and $q$ on a same discrete space, the relative entropy of $p$ with respect to $q$ is
\begin{equation}\label{eq:defS}
\Srel pq=\sum_{x} p(x)\log \frac{p(x)}{q(x)},
\end{equation}
which is always non-negative, it vanishes if and only if $p=q$, while it gives $+\infty$ whenever ${\rm supp}\,p\subsetneq{\rm supp}\,q$; the usual convention $0\log (0/0)=0$ is understood. We use base 2 logarithms, so that the entropy will be measured in bits.
The interpretation in terms of information loss can be understood in the framework of data compression theory \cite[Theor.\ 5.4.3]{CovT06}: $\Srel pq$ is the increase in the expected description length of a Shannon code when the latter is optimized assuming that a random signal has distribution $q$, but actually the true distribution is $p$.

Within this informational approach, a Heisenberg-type uncertainty relation for $\Ao$ and $\Bo$ is a trade-off relation between $\Srel{\Ao^\rho}{\Mo_{[1]}^\rho}$ and $\Srel{\Bo^\rho}{\Mo^\rho_{[2]}}$ such that no $\Mo$ can make them both too small. Any trade-off of this kind is a consequence of the incompatibility of $\Ao$ and $\Bo$, and so it reveals their quantum nature.

Depending on the framework and on the aim, one can look for relations in a specific state $\rho$ (state dependent entropic MURs), or one can first process these two relative entropies into a state-independent quantification of the quality of the approximating device $\Mo$ and then look for nontrivial constraints
(state independent entropic MURs).

In both cases, we sum the two information losses and then look for some positive lower bound for such a sum. Thus, we introduce the state-dependent \emph{error function}
\begin{equation}\label{statedependentdivergence}
\sddiv{\Ao,\Bo}{\Mo}(\rho)=\Srel{\Ao^\rho}{ \Mo^\rho_{[1]}}+ \Srel{\Bo^\rho}{\Mo^\rho_{[2]}}= \Srel{\Ao^\rho\otimes \Bo^\rho}{ \Mo^\rho_{[1]}\otimes \Mo^\rho_{[2]}}.
\end{equation}
The last equality gives the informational interpretation of the error function: it is the increase in expected length of a Shannon code describing the random signal $(\Ao,\Bo)$, if the signal has distribution $\Ao^\rho\otimes\Bo^\rho$, but it is coded using the wrong distribution $\Mo^\rho_{[1]}\otimes \Mo^\rho_{[2]}$. Thus, it is the total loss of information due to the approximations $\Ao^\rho\simeq\Mo^\rho_{[1]}$ and $\Bo^\rho\simeq\Mo^\rho_{[2]}$. Note that $\Ao^\rho\otimes \Bo^\rho$ is the distribution of a measurement of  $\Ao$ and $\Bo$ (possibly incompatible observables) in two independent preparations of the same state $\rho$; coherently, when $\Mo^\rho$ is available in place of $\Ao^\rho\otimes \Bo^\rho$, we take into account this independence and, so, first we compute the marginals $\Mo^\rho_{[1]}$ and $\Mo^\rho_{[2]}$, and then we optimize the Shannon code according to $\Mo^\rho_{[1]}\otimes \Mo^\rho_{[2]}$ and not to $\Mo^\rho$. Indeed, if $\Ao$ and $\Bo$ are incompatible, correlations between their values in a single experiment simply do not exist, and so we do not care about correlations emerging from joint measurements of
$\Mo_{[1]}$ and $\Mo_{[2]}$. In other words, different approximate joint measurements $\Mo$ and $\Mo'$ with the same marginals $\Mo_{[1]}=\Mo'_{[1]}$ and $\Mo_{[2]}=\Mo'_{[2]}$ give the same error function.

Given the target observables $\Ao$ and $\Bo$, our aim is to minimize $\sddiv{\Ao,\Bo}{\Mo}(\rho)$: we want to find out the minimal amount of information that is unavoidably lost in an approximate joint measurement (entropic MUR), and, possibly, we want to characterize the optimal devices $\Mo$ minimizing such information loss.

Similarly, if we approximate three sharp spin components $\Ao,\Bo,\Co$ with a joint measurement $\Mo$ of marginals $\Mo_{[1]},
\Mo_{[2]},\Mo_{[3]}$, then the error function is
\begin{equation}\label{statedependentdivergence3}
\sddiv{\Ao,\Bo,\Co}{\Mo}(\rho)=\Srel{\Ao^\rho}{ \Mo^\rho_{[1]}}+ \Srel{\Bo^\rho}{\Mo^\rho_{[2]}}+ \Srel{\Co^\rho}{\Mo^\rho_{[3]}}.
\end{equation}

\section{State dependent Entropic MURs}\label{sec3}
Any study of state dependent MURs between $\Ao$ and $\Bo$ must be preceded by a consideration: given a state $\rho_*$, we can always find a bi-observable $\Mo_*$ such that $\sddiv{\Ao,\Bo}{\Mo_*}(\rho_*)=0$. It suffices to take $\Mo_*=(\Ao^{\rho_*}\otimes\Bo^{\rho_*})\,\id$, the trivial bi-observable giving an output distributed as $\Ao^{\rho_*}\otimes\Bo^{\rho_*}$, independently of the system state $\rho$. For this reason it make sense to look for state-dependent MURs only if some criterion reduces the class of the approximate joint measurements $\Mo$. A typical criterion is to allow only for POVMs $\Mo$ sharing the same covariance properties of the target observables $\Ao$ and $\Bo$.

For a spin-1/2 quantum system, let us start with maximally incompatible $\Ao$ and $\Bo$, i.e.\ the case of orthogonal $\vec a$ and $\vec b$. To fix the notations, let us set
\begin{equation}\label{def:XY}
\Ao(x)=\Xo(x) = \frac{1}{2}\,(\id + x\sigma_1), \qquad \Bo(y)=\Yo(y) = \frac{1}{2}\,(\id + y\sigma_2).
\end{equation}
Then, their symmetry properties \cite{BGT18} with respect to the order 8 dihedral group $D_4$, generated by the $180^\circ$ rotations around the $x$-axis and around the bisector of the first and third quadrant, lead us to consider $D_4$-covariant bi-observable on $\{-1,+1\}^2$, that is \cite{BGT18}
\begin{equation}\label{gencovD4}
\Mo(x,y)= \frac{1}{4}\Big[\id + c\left(x\sigma_1 + y\sigma_2\right) \Big] , \qquad \abs{c}\leq 1/\sqrt 2,
\end{equation}
Then we have:
\begin{itemize}
\item for every state $\rho$ there is a $D_4$-covariant bi-observable that minimizes $\sddiv{\Xo,\Yo}{\Mo}(\rho)$,
\begin{equation}\label{optimalM2orth}
\Mo_0(x,y)= \frac{1}{4}\left(\id + \frac{x}{\sqrt{2}}\,\sigma_1 + \frac{y}{\sqrt{2}}\,\sigma_2\right),
\end{equation}
\item the optimal approximate joint measurement $\Mo_0$ is independent of $\rho$,
\item the marginals of $\Mo_0$ are equally noisy versions of the target observables,
$$
\Mo_{0\,[1]}=\frac{1}{\sqrt2}\,\Xo+\left(1-\frac{1}{\sqrt2}\right)\frac{\id}{2},\qquad
\Mo_{0\,[2]}=\frac{1}{\sqrt2}\,\Yo+\left(1-\frac{1}{\sqrt2}\right)\frac{\id}{2},
$$
\item $\Mo_0$ provides a lower bound for the error function:
\begin{multline}\label{sd2orthmur}
\sddiv{\Xo,\Yo}{\Mo}(\rho)\geq\sddiv{\Xo,\Yo}{\Mo_0}(\rho)
=\frac{1+r_1}{2}\,\log\frac{1+r_1}{1+r_1/\sqrt2}
+\frac{1-r_1}{2}\,\log\frac{1-r_1}{1-r_1/\sqrt2}\\
+\frac{1+r_2}{2}\,\log\frac{1+r_2}{1+r_2/\sqrt2}
+\frac{1-r_2}{2}\,\log\frac{1-r_2}{1-r_2/\sqrt2}
\end{multline}
for every state $\rho$ and every $D_4$-covariant approximate joint measurement $\Mo$,
\item if the state vector $\vec r$ is not perpendicular to the plane $xy$, then the lower bound \eqref{sd2orthmur} is strictly positive and $\Mo_0$ is the unique optimal $D_4$-covariant approximate joint measurement,
\item if $\vec r$ is perpendicular to the plane $xy$, then the lower bound \eqref{sd2orthmur} vanishes and every $D_4$-covariant  bi-observable $\Mo$ is optimal.
\end{itemize}
The inequality \eqref{sd2orthmur} is our state-dependent entropic MUR for two orthogonal spin components. Depending on the state $\rho$ it can be significant or trivial. Changing the point of view, when $\rho$ is given, the MUR \eqref{sd2orthmur} indicates an unavoidable information loss if and only if we perform a covariant approximate joint measurement of two spin components in a plane non perpendicular to $\vec r$. Otherwise $\Ao^\rho$, $\Mo_{[1]}^\rho$, $\Bo^\rho$ and $\Mo_{[2]}^\rho$ are all equal to the uniform distribution on $\{-1,+1\}$ and no criterion based on the output distributions can detect the approximation.

A similar result holds for approximate joint measurements of three maximally incompatible spin-1/2 components, say
\begin{equation}\label{def:XYZ}
\Xo(x) = \frac{1}{2}\,(\id + x\sigma_1), \qquad \Yo(y) = \frac{1}{2}\,(\id + y\sigma_2), \qquad \Zo(z) = \frac{1}{2}\,(\id + z\sigma_3).
\end{equation}
Then, their symmetry properties \cite{BGT18} with respect to the order 24 octahedron group $O$, generated by the $90^\circ$ rotations around the three coordinate axes, lead us to consider $O$-covariant tri-observable on $\{-1,+1\}^3$, that is \cite{BGT18}
\begin{equation}\label{gencovO}
\Mo(x,y,z)= \frac{1}{8}\Big[\id + c\left(x\sigma_1 + y\sigma_2 + z\sigma_3\right) \Big] , \qquad \abs{c}\leq 1/\sqrt 3.
\end{equation}
Then we have:
\begin{itemize}
\item for every $\rho$ there is a $O$-covariant tri-observable that minimizes $\sddiv{\Xo,\Yo,\Zo}{\Mo}(\rho)$,
\begin{equation}\label{optimalM3orth}
\Mo_0(x,y,z)= \frac{1}{8}\left(\id + \frac{x}{\sqrt{3}}\,\sigma_1 + \frac{y}{\sqrt{3}}\,\sigma_2 + \frac{z}{\sqrt{3}}\,\sigma_3\right),
\end{equation}
\item the optimal approximate joint measurement $\Mo_0$ is independent of $\rho$,
\item the marginals of $\Mo_0$ are equally noisy versions of the target observables,
\begin{equation}\begin{split}
\Mo_{0\,[1]}=\frac{1}{\sqrt3}\,\Xo+&\left(1-\frac{1}{\sqrt3}\right)\frac{\id}{2},\qquad
\Mo_{0\,[2]}=\frac{1}{\sqrt3}\,\Yo+\left(1-\frac{1}{\sqrt3}\right)\frac{\id}{2},\\
&\Mo_{0\,[3]}=\frac{1}{\sqrt3}\,\Zo+\left(1-\frac{1}{\sqrt3}\right)\frac{\id}{2},
\end{split}
\end{equation}
\item $\Mo_0$ provides a lower bound for the error function:
\begin{multline}\label{sd3orthmur}
\sddiv{\Xo,\Yo,\Zo}{\Mo}(\rho)\geq\sddiv{\Xo,\Yo,\Zo}{\Mo_0}(\rho)\\
=\sum_{k=1}^3 \left\{\frac{1+r_k}{2}\,\log \frac {1+r_k} {1+r_k/\sqrt 2}+ \frac{1-r_k}{2}\,\log \frac {1-r_k} {1-r_k/\sqrt 2}\right\}
\end{multline}
for every state $\rho$ and every $O$-covariant approximate joint measurement $\Mo$,
\item for every $\rho\neq\frac{\id}{2}$ the lower bound \eqref{sd3orthmur} is strictly positive and $\Mo_0$ is the unique optimal $O$-covariant approximate joint measurement,
\item for $\rho=\frac{\id}{2}$ the lower bound \eqref{sd3orthmur} vanishes and every $O$-covariant tri-observable $\Mo$ is optimal.
\end{itemize}
The inequality \eqref{sd3orthmur} is our state-dependent entropic MUR for three orthogonal spin components. Differently from the two components case, the lower bound now vanishes only for the maximally chaotic state.

\section{State independent Entropic MURs}\label{sec4}
In this section, every POVM on $\{-1,+1\}^2$ will be considered as a possible approximate joint measurement of the target observables $\Ao$ and $\Bo$.

If one is interested in the quality of a joint measurement device $\Mo$, whatever the state $\rho$ of the observed system, then a state-independent quantification of the approximation error is needed.

Note that the minimum error is not significative, as it can be null for an approximate joint measurement $\Mo$ of incompatible $\Ao$ and $\Bo$ (as we have seen in the previous section). That is, it can be null for some $\rho$ even if $\Mo_{[1]}\neq\Ao$ or $\Mo_{[2]}\neq\Bo$.

Then typical approaches consist in computing some mean error or the worst error. Following \cite{BusLW14}, we maximize our error function.
The \emph{divergence} of $\Mo$ from $(\Ao,\Bo)$ is the worst-case information loss
\begin{equation}\label{eqdef:D}
\Div{\Ao,\Bo}{\Mo}= \sup_{\rho} \sddiv{\Ao,\Bo}{\Mo}(\rho).
\end{equation}
Note that $\Div{\Ao,\Bo}{\Mo}$ takes into account any possible balancing and compensation between the information losses in the two approximations $\Ao^\rho\simeq\Mo^\rho_{[1]}$ and $\Bo^\rho\simeq\Mo^\rho_{[2]}$. It is always non-negative and it vanishes exactly when $\Ao$ and $\Bo$ are compatible and $\Mo$ is one of their joint measurements. Given $\Ao$ and $\Bo$, our aim is to make the divergence $\Div{\Ao,\Bo}{\Mo}$ as small as possible.

Again, explicit computations can be done in the case of maximally incompatible $\Ao$ and $\Bo$, i.e. the observables $\Xo$ and $\Yo$ \eqref{def:XY}. Then \cite{BGT18}
\begin{itemize}
\item there is a unique optimal approximate joint measurement of $\Xo$ and $\Yo$, that is the covariant bi-observable $\Mo_0$ \eqref{optimalM2orth},
\item if $\rho_e$ is the projection on any eigenvector of $\sigma_1$ or $\sigma_2$, then
\begin{equation}\label{corth}
\Div{\Xo,\Yo}{\Mo_0}=\sddiv{\Xo,\Yo}{\Mo_0}(\rho_{e}) =\log\left[2\left(2-\sqrt 2\right)\right],
\end{equation}
\item $\Mo_0$ provides a lower bound for the divergence:
\begin{equation}\label{si2orthmur}
\Div{\Xo,\Yo}{\Mo}\geq\log \left[2\left(2-\sqrt 2\right)\right]
\end{equation}
for every approximate joint measurement $\Mo$.
\end{itemize}
The inequality \eqref{si2orthmur} is our state-independent entropic MUR, which can be stated also as a statement about the total loss of information that occurs in a single preparation of the system: for every approximate joint measurement $\Mo$ of $\Xo$ and $\Yo$, there exists a state $\rho$ such that
\begin{equation}\label{MUR2}
\Srel{\Xo^{\rho}}{\Mo^{\rho}_{[1]}}+\Srel{\Yo^{\rho}}{\Mo^{\rho}_{[2]}}\geq \log\left[2\left(2-\sqrt 2\right)\right].
\end{equation}
So, in an approximate joint measurement of $\Xo$ and $\Yo$, the total loss of information can not be arbitrarily reduced by the choice of the device $\Mo$: it depends on the state $\rho$, but potentially it can always be as large as $\log\left[2\left(2-\sqrt 2\right)\right]$.

A similar result holds for approximate joint measurements of three maximally incompatible spin-1/2 components, i.e.\ the observables $\Xo$, $\Yo$, $\Zo$ \eqref{def:XYZ}. The definition of $\Div{\Xo,\Yo,\Zo}{\Mo}$ is obvious. Then \cite{BGT18}
\begin{itemize}
\item there is a unique optimal $O$-covariant approximate joint measurement of $\Xo$, $\Yo$, $\Zo$, that is the tri-observable $\Mo_0$ \eqref{optimalM3orth},
\item if $\rho_e$ is the projection on any eigenvector of $\sigma_1$ or $\sigma_2$ or $\sigma_3$, then
\begin{equation}\label{corth3}
\Div{\Xo,\Yo,\Zo}{\Mo_0}=\sddiv{\Xo,\Yo,\Zo}{\Mo_0}(\rho_{e})=\log\left(3-\sqrt3\right),
\end{equation}
\item $\Mo_0$ provides a lower bound for the divergence:
\begin{equation}\label{si3orthmur}
\Div{\Xo,\Yo,\Zo}{\Mo}\geq\log\left(3-\sqrt3\right)
\end{equation}
for every approximate joint measurement $\Mo$.
\end{itemize}
Note that, differently from the two components case, the uniqueness of the optimal approximate joint measurement is stated only in the class of $O$-covariant bi-observables; indeed, a non-covariant tri-observable with the same marginals as $\Mo_0$ can be easily exhibited \cite{HeJiReZi10}.

For two general components $\Ao$ and $\Bo$, let $0\leq\alpha\leq\pi$ be the angle between $\vec{a}$ and $\vec{b}$. We introduce the \emph{entropic incompatibility degree} of the observables $\Ao$ and $\Bo$,
\begin{equation}\label{def:inc}
\Icomp(\alpha)=\inf_{\Mo} \Div{\Ao,\Bo}{\Mo} = \inf_{\Mo}\sup_{\rho} \sddiv{\Ao,\Bo}{\Mo}(\rho).
\end{equation}
Indeed, the infimum \eqref{def:inc} depends only on the angle $\alpha$. Since $\Icomp(\alpha)$ vanishes if and only if $\alpha=0,\pi$, that is for compatible $\Ao$ and $\Bo$, we have the state independent entropic MUR:
\begin{equation}\label{MURalpha}
\Div{\Ao,\Bo}{\Mo}\geq \Icomp(\alpha)
\end{equation}
for every approximate joint measurement $\Mo$. By construction, the lower bound $\Icomp(\alpha)$ is tight. So its name is justified.

In this general case, we choose the coordinate axes in such a way that the bisector of $\alpha$ coincides with the bisector of the first quadrant. Then the symmetry group of $\Ao$ and $\Bo$ is the order 4 dihedral group $D_2$ generated by the $180^\circ$ rotations around the two bisectors of the quadrants.

We do not know if the optimal approximate joint measurement $\Mo$ of $\Ao$ and $\Bo$ is unique and if it needs to be $D_2$-covariant. Anyway, we know that there is at least an optimal one in the family, consisting of $D_2$-covariant bi-observables,
\begin{equation}\label{Mgamma}
\Mo_\gamma(x,y) = \frac{1}{4}\left[ \left(1 + \gamma xy\right)\id + \frac{1}{\sqrt{2}} \left(x\sigma_1 + y\sigma_2\right) + \frac{\gamma}{\sqrt{2}} \left(y\sigma_1 + x\sigma_2\right)\right], \qquad \abs{\gamma}\leq 1.
\end{equation}
Then numerical computations give the following graph of $\Icomp(\alpha)$ as a function of the angle $\alpha$.
Just as one would expect, the entropic incompatibility degree $\Icomp(\alpha)$ is null for compatible observables, $\alpha=0$, and then continuously increases to its maximum value given by maximally incompatible observables, $\alpha=\pi/2$. Then the graph is symmetric, as it should be, with respect to $\alpha=\pi/2$.
\begin{SCfigure}[50][h!]
\includegraphics*[width=9cm,height=5cm]
{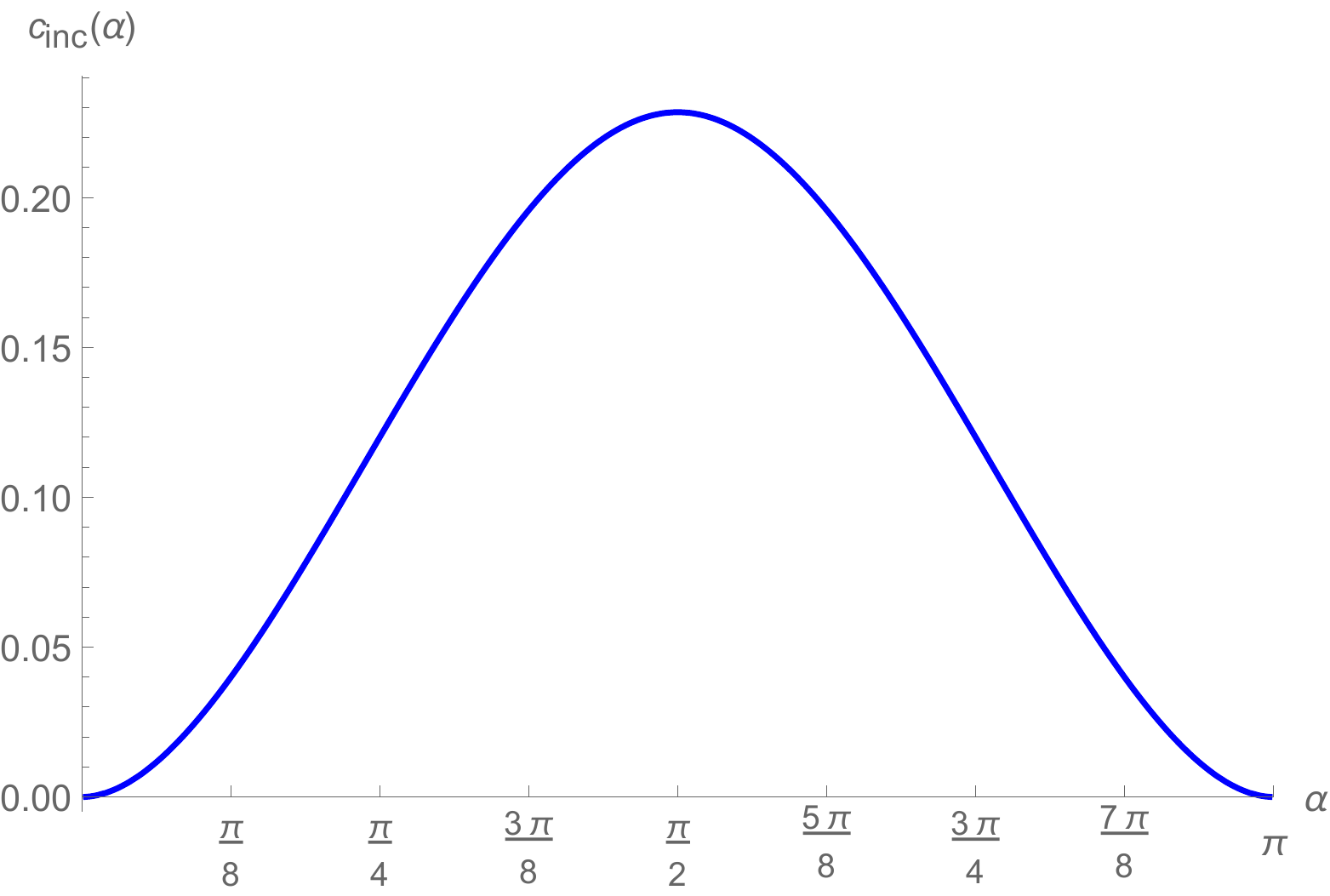}
\caption{Numerical evaluations of $\Icomp(\alpha)$ as a function of $\alpha$, from \eqref{def:inc}, \eqref{Mgamma}: $\Icomp(\alpha)= \inf_{\gamma}\sup_{\rho} \sddiv{\Ao,\Bo}{\Mo_\gamma}(\rho)$.}\label{mod}
\end{SCfigure}

Numerical computations also provide an optimal $\gamma_0$ for every chosen $\alpha$ and it turns out that, differently from the orthogonal cases, for a general $\alpha$ the marginals of the optimal approximate joint measurement $\Mo_{\gamma_0}$ are not noisy versions of the target observables $\Ao$ and $\Bo$ \cite{BGT18}; similar results hold also for different quantifications of the uncertainty \cite{BusH08,HSR08,BullB18}. This shows that it is not always possible to reduce the study of approximate joint measurements of sharp observables to the study of their noisy versions.

\section{Entropic MURs for infinite components}\label{sec5}
Now we consider an approximate joint measurement of all the spin components $\Ao$, $|\vec a|=1$. Since $\vec a$ is not fixed anymore, we change notation and, for every $\vec a$, we denote by $\Ao_{\vec a}$ the corresponding observable.

The following is inspired by \cite{DamSW15}, where the same problem is considered for a generic spin $s$, but the approximation error is quantified by Wasserstein distances between target and approximating distributions, and uncertainty regions are introduced.

First we consider a POVM observable $\Fo$ with output $\vec\xi\in\mathbb{R}^3$, $|\vec\xi|=1$. From this, by post-processing we derive our approximate joint measurement $\Mo$ of all spin components $\Ao_{\vec a}$: if $\vec\xi$ is the output of $\Fo$, then the output $x=\pm1$ of the approximate measurement $\Mo_{[\vec a]}$ along $\vec a$ is the signum the $\vec a$-component $\vec\xi\cdot\vec a$. Thus the marginals $\Mo_{[\vec a]}$ are all compatible by construction and, of course, their outputs are all ``classically coherent''.

Just as in \cite{DamSW15}, we want $\Fo$ to represent an angular momentum and, so, we consider only $SO(3)$-covariant $\Fo$, that is (Eq.\ (108) of \cite{DamSW15})
\begin{equation}\label{Fep}
\Fo_\epsilon(\rmd \theta,\rmd \phi)=(1-\epsilon)\Fo_+(\rmd \theta,\rmd \phi)+\epsilon \Fo_-(\rmd \theta,\rmd \phi), \qquad \epsilon\in [0,1],
\end{equation}
\begin{equation}\label{Fpm}
\Fo_\pm(\rmd \theta,\rmd \phi):=\frac{\sin \theta \rmd \theta\rmd\phi}{4\pi}\left[\openone\pm \sin \theta\left( \cos \phi \,\sigma_1 + \sin \phi\, \sigma_2\right)\pm \cos \theta \,\sigma_3\right].
\end{equation}
We are using spherical coordinates with the usual conventions $\vec\xi=(\sin\theta\cos\phi,\sin\theta\sin\phi,\cos\theta)$, $0\leq\theta\leq\pi$, $0\leq\phi<2\pi$; the choice of the $z$-axis is arbitrary. Note that $\frac{\sin \theta}{4\pi}\,\rmd \theta\rmd \phi$ is the uniform distribution on the unit sphere.

Then, the marginals of the corresponding approximate joint measurement $\Mo_\epsilon$ turn out to be
\begin{equation}\label{marginal}
\Mo_{\epsilon\,[\vec a]}(x)=\frac 12 \left(\id+\frac{1-2\epsilon}2\, x \vec a \cdot \vec \sigma \right), \qquad x=\pm 1.
\end{equation}

In this case we can not sum infinite relative entropies, but we can slightly modify our approach and, instead of computing the total information loss, we consider the mean information loss. Thus, our \emph{error function} now is
\begin{equation}\label{infiniteerrorfz}
\overline{S}[\Acal\|\Mo_\epsilon](\rho)=\int_0^\pi\rmd \theta \,\frac {\sin \theta}{4\pi} \int_0^{2\pi}\rmd \phi\,S\big(\Ao_{\vec a}^\rho\big\| \Mo_{\epsilon\,[\vec a]}^\rho\big), \qquad \vec a=\begin{pmatrix} \sin \theta \cos \phi \\ \sin \theta \sin \phi \\ \cos \theta\end{pmatrix},
\end{equation}
where $\Acal=\{\Ao_{\vec a}\}_{\vec a}$ denotes the collection of all the possible spin components. Then
\begin{itemize}
\item for every state $\rho$ there is a $SO(3)$-covariant observable $\Fo_0$ such that the corresponding $\Mo_0$ minimizes $\overline{S}[\Acal\|\Mo_\epsilon](\rho)$,
\begin{equation}\label{optimalF}
\Fo_0(\rmd \theta,\, \rmd \phi)= \Fo_+(\rmd \theta,\, \rmd \phi)=\Big\{\openone+ \sin \theta\left( \cos \phi \,\sigma_1 + \sin \phi\, \sigma_2\right)+ \cos \theta \,\sigma_3\Big\}\frac{\sin \theta}{4\pi}\,\rmd \theta\rmd \phi,
\end{equation}
\item the optimal observable $\Fo_0$ is independent of $\rho$,
\item the marginals of $\Mo_0$ are equally noisy versions of the target observables,
$$
\Mo_{0\,[\vec a]}=\frac{1}{2}\,\Ao_{\vec a}+\frac{1}{2}\,\frac{\id}{2},
$$
\item $\Fo_0$ provides a lower bound for the error function:
\begin{multline}\label{infinitecompmur}
\overline{S}[\Acal\|\Mo_\epsilon](\rho)\geq\overline{S}[\Acal\|\Mo_0](\rho)\\
=\frac{(1+r)^2}{4r}\,\log \frac{2(1+r)}{2+r} + \frac{(1-r)^2}{4r}\,\log \frac{2-r}{2(1-r)} + \frac{1}{4r}\,\log \frac{2+r}{2-r}-\frac{\log\rme}{2}
\end{multline}
for every $\rho$ and every approximate joint measurement $\Mo_\epsilon$ coming from a $SO(3)$-covariant $\Fo_\epsilon$,
\item for every $\rho\neq\frac{\id}{2}$ the lower bound \eqref{infinitecompmur} is strictly positive and $\Fo_0$ is the unique optimal $SO(3)$-covariant observable,
\item for $\rho=\frac{\id}{2}$ the lower bound \eqref{infinitecompmur} vanishes and every $SO(3)$-covariant observable $\Fo_\epsilon$ is optimal.
\end{itemize}
The inequality \eqref{infinitecompmur} is our state-dependent entropic MUR. The lower bound is always non negative and it vanishes only for the maximally chaotic state.

Let us note that \cite{DamSW15} characterizes the error of the approximating device with respect to the target reference by the worst direction $\vec a$. Similarly, we could find an alternative entropic MUR by studying $\max_{\vec a}S\big(\Ao_{\vec a}^\rho\big\| \Mo_{\epsilon\,[\vec a]}^\rho\big)$, but we prefer our error function because it allows a direct comparison with the previous sections.

In this framework, we introduce the \emph{divergence} of $\Mo_\epsilon$ from the collection $\Acal$ of all the spin components as the worst-case mean information loss,
\begin{equation}\label{eqdef:mD}
\Divm{\Acal}{\Mo_\epsilon}= \sup_{\rho} \overline{S}[\Acal\|\Mo_\epsilon](\rho).
\end{equation}
Then
\begin{itemize}
\item $\Divm{\Acal}{\Mo_\epsilon}=\overline{S}[\Acal\|\Mo_\epsilon](\rho_*)$ for every \emph{pure} state $\rho_*$,
\item there is a unique $SO(3)$-covariant observable $\Fo_\epsilon$ giving an optimal approximate joint measurement $\Mo$ of all spin components $\Ao_{\vec a}$, that is $\Fo_0$ \eqref{optimalF},
\item  the corresponding measurement $\Mo_0$ provides a lower bound for the divergence: for every approximate joint measurement $\Mo_\epsilon$ coming from a $SO(3)$-covariant observable $\Fo_\epsilon$ we have
\begin{equation}\label{siinfmur}
\Divm{\Acal}{\Mo_\epsilon}\geq \Divm{\Acal}{\Mo_0}=\frac 3 4 \,\log\frac 4 3 -\frac{\log \rme -1}2.
\end{equation}

\end{itemize}
The inequality \eqref{siinfmur} is our state independent entropic MUR, which now can be formulated also as a statement about the mean loss of information that occurs in every \emph{pure} preparation of the system: for every approximate joint measurement $\Mo_\epsilon$ of all the spin components,
\begin{equation}
\overline{S}[\Acal\|\Mo_\epsilon](\rho)\geq\frac 3 4 \,\log\frac 4 3 -\frac{\log \rme -1}2,
\end{equation}
for all the pure states $\rho$.

In order to compare the lower bound in the previous MURs \eqref{si2orthmur} and \eqref{si3orthmur} with \eqref{siinfmur}, we divide each member in \eqref{si2orthmur} and \eqref{si3orthmur} by the number of the target observables, so that every time the mean information loss is considered. The lower bounds in the three resulting MURs are:
\[
\overline\Icomp(\Xo,\Yo)=\frac 12\,\log \left[2\left(2-\sqrt 2\right)\right] \simeq 0.114223
\]
\[
\overline \Icomp(\Xo,\Yo,\Zo)=\frac 13\log \left(3-\sqrt 3\right)\simeq 0.114166
\]
\[
\overline\Icomp(\Acal)=\frac 3 4 \,\log\frac 4 3 -\frac{\log \rme -1}2=2-\frac 34\,\log 3 -\frac 12\, \log \rme \simeq 0.0899306040
\]
so that
$$
\overline\Icomp(\Acal)<\overline\Icomp(\Xo,\Yo,\Zo)<\overline\Icomp(\Xo,\Yo),
$$
which shows that the higher mean loss of information is in the first step: to make compatible two orthogonal components.

\section{Proofs}\label{sec6}

The results of Sections \ref{sec3} and \ref{sec5} are new and they are proved in the following.

\paragraph{Proofs of Section 3.}
The results for the two orthogonal spin components $\Xo$ and $\Yo$ come from the direct computation of the error function for $D_4$-covariant bi-observables $\Mo$ \eqref{gencovD4}:
$$
\sddiv{\Xo,\Yo}{\Mo}(\rho)=\sum_{k=1}^2 \left\{\frac{1+r_k}{2}\,\log \frac {1+r_k} {1+cr_k}+ \frac{1-r_k}{2}\,\log \frac {1-r_k} {1-cr_k}\right\}=\sum_{k=1}^2s_k(c).
$$
If $\vec r$ is perpendicular to the plane $xy$, then $r_1=r_2=0$ and the error function $\sddiv{\Xo,\Yo}{\Mo}(\rho)$ is null for every $c$, so that \eqref{optimalM2orth} trivially holds. On the other hand, for every $r_k\neq0$, the function $s_k(c)$ is decreasing in $c$ because of general properties of relative entropy: the function is convex in $c\in[-1,1]$, vanishes in $c=1$ and is positive elsewhere. As we can allow only for $\abs{c}\leq 1/\sqrt 2$, the optimal observable $\Mo_0$ is given by $c=1/\sqrt 2$, that is \eqref{optimalM2orth}.
In both cases, \eqref{optimalM2orth} implies all the other results for $\Xo$ and $\Yo$.

Similarly, the results for the three orthogonal spin components $\Xo$, $\Yo$ and $\Zo$ come from the direct computation of the error function for $O$-covariant bi-observables $\Mo$ \eqref{gencovO}:
$$
\sddiv{\Xo,\Yo,\Zo}{\Mo}(\rho)=\sum_{k=1}^3 \left\{\frac{1+r_k}{2}\,\log \frac {1+r_k} {1+cr_k}+ \frac{1-r_k}{2}\,\log \frac {1-r_k} {1-cr_k}\right\}.
$$

\paragraph{Proofs of Section 5.}
Given the $SO(3)$-covariant observable $\Fo_\epsilon$ \eqref{Fep}, the marginals \eqref{marginal} come from direct computation. Then the error function $\overline{S}[\Acal\|\Mo_\epsilon](\rho)$ \eqref{infiniteerrorfz} can be explicitly computed and its monotone properties with respect to $r$ and $\epsilon$ can be studied.

By rotation invariance we can take the $z$-axis parallel to $\vec r$; then $\vec a\cdot \vec r=r\cos \theta$, and the relative entropy of each approximation is
$$
S\big(\Ao_{\vec a}^\rho\big\| \Mo_{\epsilon\,[\vec a]}^\rho\big)=\sum_{x=\pm1}\frac{1+xr\cos\theta}{2}\,\log\frac{1+xr\cos\theta}{1+\lambda xr\cos\theta}\,, \qquad \lambda=\frac{1-2\epsilon}{2}\in[-1/2,1/2].
$$
This quantity is decreasing in $\lambda\in[-1/2,1/2]$ if $r\cos\theta\neq0$, as it is convex in $\lambda\in[-1,1]$ and it vanishes if and only if $\lambda=1$, and it is increasing in $r\in[0,1]$, as it is convex in $r\in[0,1]$ and it vanishes if and only if $r=0$.
Then the error function \eqref{infiniteerrorfz} becomes
$$
\overline{S}[\Acal\|\Mo_\epsilon](\rho)=\frac{1}{4}\sum_{x=\pm1}\int_0^\pi  \sin\theta\left(1+xr\cos\theta\right)\log\frac{1+xr\cos\theta}{1+\lambda xr\cos\theta}\,\rmd \theta\,,
$$
which is increasing in $r\in[0,1]$ if $\lambda \neq 0$, and, if $r\neq0$, it is decreasing in $\lambda\in[-1/2,1/2]$ and increasing in $\epsilon\in[0,1]$.
By the changes of variable $z=xr \cos \theta$ we get
\[
\overline{S}[\Acal\|\Mo_\epsilon](\rho)
= \frac {\log \rme} {2r}\int_{-r}^r (1+z)\ln \frac{1+z}{1+\lambda z}\,\rmd z ,
\]
and using
\[
\frac 12 \,\frac{\rmd \ }{\rmd z}\left\{ (1+z)^2\ln\frac{1+z}{1+\lambda z}+\frac{1-\lambda}{\lambda}\left[\frac{1-\lambda}{\lambda}\,\ln \left(1+\lambda z\right)-z\right]\right\}=(1+z)\ln\frac{1+z}{1+\lambda z}\,,
\]
finally we get
\begin{multline}\label{erf}
\overline{S}[\Acal\|\Mo_\epsilon](\rho)= \frac{(1+r)^2}{4r}\,\log \frac{2(1+r)}{2+(1-2\epsilon)r} - \frac{(1-r)^2}{4r}\,\log \frac{2(1-r)}{2-(1-2\epsilon)r} \\ {}+ \frac{(1+2\epsilon)^2}{4(1-2\epsilon)^2r}\,\log \frac{2+(1-2\epsilon)r}{2-(1-2\epsilon)r}-\frac{1+2\epsilon}{2(1-2\epsilon)}\,\log\rme.
\end{multline}
Equality \eqref{erf} and the monotonicity properties with respect to $r$ and $\epsilon$ imply all the other results.

\end{document}